\documentclass[runningheads]{llncs}
\usepackage{amsfonts}
\usepackage{float}
\usepackage[dvips]{graphicx}
\usepackage{array}
\usepackage{multirow}
\usepackage{alltt}
\usepackage{alltt}

\usepackage{macrowlpe07}

\begin{document}

%\selectlanguage{francais}
%\creationEntete

\mainmatter \title{Observational Semantics of the Prolog Resolution Box Model}
%\mainmatter \title{S\'emantique observationnelle du modèle des boîtes pour la r\'esolution Prolog}
%
\titlerunning{The Resolution Box Model}
\author{Pierre Deransart\inst{1} \and Mireille Ducass\'e\inst{2} \and G\'erard Ferrand\inst{3}}
\authorrunning{Deransart \& al}
\institute{{\sc Inria} Rocquencourt, BP 105, 78153 Le Chesnay Cedex, France\\
\email{Pierre.Deransart@inria.fr}
\and {\sc INSA}, Dept Info, 20, av des Buttes de Coesmes, 35043 Rennes Cedex, France\\
\email{Mireille.Ducasse@irisa.fr}
\and {\sc LIFO}, BP 6759, 45067 Orl\'eans Cedex 2, France\\
\email{Gerard.Ferrand@lifo.univ-orleans.fr}
}

\maketitle

\begin{abstract}
  This paper specifies an observational semantics and gives an
  original presentation of the Byrd box model. The approach accounts
  for the semantics of Prolog tracers independently of a particular
  Prolog implementation.

  Prolog traces are, in general, considered as rather obscure and difficult
  to use. The proposed formal presentation of its trace constitutes a
  simple and pedagogical approach for teaching Prolog or for
  implementing Prolog tracers. It is a form of declarative
  specification for the tracers.

  The trace model introduced here is only one example to
  illustrate general problems relating to tracers and observing
  processes.  Observing processes know, from observed processes, only
  their traces. The issue is then to be able to reconstitute, by the
  sole analysis of the trace, part of the behaviour of the observed process, and if
  possible, without any loss of information.

 As a matter of fact, our approach highlights qualities of the Prolog resolution box model which made its
 success, but also its insufficiencies.

\end{abstract}

\section{Introduction}
\label{intro}
This paper presents a Prolog trace model, often called \emph{Byrd box
model}, in an original way, based on the concept of Observational
Semantics (OS). This semantics was introduced in~\cite{ercimlnai} in
order to account for the semantics of tracers independently of the
semantics of the traced process. 

The objective of this paper is to illustrate the merits of an
observational semantics with a simple but non trivial example. The
result is an original semantics of the Prolog trace as usually
implemented, but without either taking into account any particular
implementation or describing the totality of the resolution
process. Such a semantics also constitutes a form of formal
specification of Prolog tracers and makes it possible to easily
understand some of their essential properties.

%Understanding a trace is somehow equivalent to trying to retrieve the operation of the traced process from a known initial state and a succession of trace events. This approach requires a sufficient, but not necessarily complete, knowledge of the operational model of the process. It also assumes that one knows how to connect trace events to this model. The trace ``faithfulness'' relative to the observed model is what we want to capture with the concepts of observational semantics, trace extraction mechanisms and the functions to rebuild the original model starting from the trace.
	
The ``box model'' was introduced for the first time by Lawrence Byrd in
1980~\cite{byrd80} to help users of the ``new'' Prolog
language~\footnote{it then refers to the implementations of
  Edinburgh~\cite{proedinb} and of Marseille~\cite {promars}} to
master the operational reading of program executions. Indeed, since
the very beginning, users have been complaining about how difficult it
is to understand control mechanisms related to the non-determinism of
the solutions.  Even if thereafter other models were adopted
with more complex strategies\footnote{the Byrd model is limited to
  the standard strategy of visiting and building a tree.}, the four
``ports'' introduced by Byrd ({\bf Call}, {\bf Exit}, {\bf Redo} and
{\bf Fail}), associated to the four corners of a box and easy to
handle in a kind of algebra of Russian headstocks, remained famous and
are more or less in all the traces of the still existing Prolog
systems.

The Byrd box model fascinates by his apparent simplicity. Often quoted
but seldomly well explained, it remains the object of sporadic but
regular publications since 1980, for example~\cite{boiz84} (1984),
\cite{ToBe93} (1993), \cite{jahier00} (2000), \cite{kulas03}
(2003). Yet, it remains difficult to explain, its various definitions
are either too abstract or drowned in a complete formalization of
Prolog operational semantics.

In this article, we propose a formal description of a variant
%% le bon mot ?
of the initial model of Byrd. The originality of this description lies
in the fact that it is formally complete, although it contains the
ingredients of the original model and it refers as little as possible
to the Prolog mechanisms of clause choice and unification.

After an introduction to the traces and the observational semantics
(Sections~\ref{introtrace} and~\ref{semobs}) which outlines the context of this study,
we present an
observational semantics which specifies the box model
(Section~\ref{soprologpur}) and the trace extraction
(Section~\ref{generation}). Finally, we give a faithful reconstruction model
(Section~\ref{reconstruction}),
%(the proof of adequacy is in appendix
which establishes a possible reading guideline for the trace, based on
the OS.
%Our approach highlights qualities of the box model which made its success. It also shows its principal defects (Section~\ref{commdisc}). This article also illustrates the potentiality of the observational approach.
This paper is based on the full report \cite{halchive} (same title and authors, in French). More details on the motivations can be found in~\cite{ercimlnai} and~\cite{pierrewlpe06}.

%On trouvera \'egalement cette \'etude plus d\'etaill\'ee dans le rapport complet de m\^emes titre et auteurs \cite{halchive}.

%---------------------------------------------------------------------

\section{Introducing  traces}
\label{introtrace}

%We first introduce the concept of full virtual trace. 

%This section gives an  outline of the context of this study. More details on the motivations can be found in~\cite{ercimlnai} and~\cite{pierrewlpe06}.

We are interested in the observation of dynamic processes starting
from the traces which they produce.

One can always consider that between an observer and an observed
phenomenon there is an object that we call \emph{trace}. The trace is
the recognized print left by a process and it is thus ``readable'' by
other processes. The observed phenomenon will be regarded here as a
closed process, its data and functions are not visible from the
outside. External processes can only know its trace. This trace is called the {\em actual trace}.
It is a sequence of trace events defined on some state.
The actual trace is thus a kind of continuous data flow produced by some process.

%The trace can be formalized by a state transition model, namely by a succession of states and a transition function formalizing the transition from a state to another. We call this semantics \emph{Observational Semantic} (OS) because it represents what one is likely to know or to describe from the process, seen from outside. The OS i characterized by the fact that each transition gives rise to a trace event. A trace can be infinite, the  number of different  transition types, however, is finite. 
%These types are traditionally called \emph{trace ports}. 
%Thus an OS is characterized by a finite set $R$ of transition rules operating on states (see next section for a concrete representation).
\vspace{1mm}
A trace can also be interpreted as the evolution of an ``abstract''
state which contains all what one can or wants to know from the process.
The sequence of the abstract states can be viewed as a more abstract but meaningful trace.
Such a trace is called the {\em full virtual trace}. 
%It can be viewed as a trace ``rebuilt'' from the actual one. 

\clearpage
%---------------------------------------------------------------------DEF
\begin{definition} [Full virtual trace]

  A full virtual trace is defined on a set of full virtual states ${\cal S}$ and a finite set of event type $R$. It is an unbounded sequence of trace events of the form {\bf $e_t: (t, a_t, S_{t+1})$} where:
\begin{itemize}
  \item $e_t$: is a unique {\bf identifier} of the event.
  \item $t$: is the {\bf chrono}, a time stamp of the event. It is an
    integer, incremented by 1 at each event.
  \item $S_{t+1} = p_{1, t+1}..., p_{n, t+1}$  is the reached state of ${\cal S}$ represented by the new values of {\bf
      parameters} $p_{1}..., p_{n}$. 
  \item $a_t$: an identifier characterizing the {\bf kind of actions} performed during
    the transition from state $S_t$  to state $S_{t+1}$, also called {\bf event type}.
\end{itemize}
%The set of event types is denoted $R$ and it is assumed to be finite.

A full virtual trace is denoted $T_v = <S_0, v^*_t> , t \geq 0$ , where $S_0$ is an initial state and $v^*_t$ a possibly empty sequence (if $t = 0$) of trace events or a sequence of length $t+1$ of the form $e_t, e_{t-1}, ..., e_0$ (if $t \geq 0$).
\end{definition}
%---------------------------------------------------------------------

%The full virtual trace represents what one wishes or what it is possible to observe of a given process. 
%As the (virtual) current state of the process is completely described in this trace, one can hope neither to produce it nor to communicate it efficiently. In practice, a kind of ``compression'' has to be carried out, and it has to be verified that the observing process can ``decompress'' accordingly. The actually produced trace will be extracted from the virtual trace and communicated in the form of an ``actual trace''.
%%%%

\begin{example}
In the Section~\ref{soprologpur} the box model is introduced with its full virtual trace. The main ``virtual objets'' described in the parameters consists of a tree whose nodes are labeled by predications and clauses. Such a tree allows to follow the evolution of the proofs during Prolog execution. Starting from a tree reduced to one root node labeled by a unique predication goal and the clauses likely used to solve it, the full virtual trace is the sequence of trees until all possible proof trees have been obtained, including failed partial trees. We do not give more detail here, since the explicit representation of a single state may be quite large and is illustrated in the Section~\ref{soprologpur}.
\end{example}

The full virtual trace may be viewed as an interpretation of the actual trace and the actual trace can be viewed as ``extracted'' from the virtual one.

%------------------------------------------------------DEF
\begin{definition} [Actual Trace, Trace Schema, Interpretation Schema]

  An actual trace is defined on a set of attribute states ${\cal A}$.
%and a finite set of event type $R$ (the same as for the full virtual trace). 
It is an unbounded sequence of trace events of the form {\bf $e_t: (t, A_{t})$}. $e_t$ and $t$ are like in the previous definition and $A_t$ is a tuple of attribute values.

An actual trace is denoted $T_w = <Q_0, w^*_t>$, with the same conventions as for the full virtual trace, where $Q_0$ is a subset of the parameters of $S_0$, the full virtual state.

Each actual trace event is derived from the ``transition'' $<S_t, S_{t+1}>$ by a function ${\cal E}$, called {\em extraction function}, and such that $A_t = {\cal E}(S_t, S_{t+1})$.

If $\forall t, A_t = (a_t, S_{t+1})$, the actual trace is a full virtual trace.

Each full virtual trace state  can be partly reconstructed from the actual trace by a function ${\cal C}$, called {\em rebuilding function}, and such that $Q_t = {\cal C}(w^*_t, Q_0)$, where $Q_t$ is a subset of $S_t$. The state domain of ${\cal C}$, denoted ${\cal Q}$, is ${\cal S}$ restricted to the parameters of ${\cal Q}$, and is called the {\em actual state domain}\footnote{${\cal A}$ and ${\cal Q}$ are different domains (except if the actual trace is the full virtual trace) and should not be confused.}.

\noindent
%$A_t$ is a finite sequence of attribute values.
%The extraction function is a family of functions defined for every transition rule $r$ of the OS:
%\\ ${\cal E} = \{{\cal
%  E}_r | r \in R \}$ such that $\forall <r,S,S'> \in OS, \ \ {\cal E}_r(S,S') = e_r$, where $e_r$ denotes the computations of the attributes.
%\noindent
%La description de la famille de fonctions ${\cal E}_r$, avec les calculs d'attributs, constitue un {\em sch\'ema de trace}.
The description of the extraction function ${\cal E}$ is called a {\em trace schema}. The description of the rebuilding  function ${\cal C}$ is called a {\em trace interpretation schema}. By definition, a rebuilding function always exists, when there may be no function of extraction.
\end{definition}

%---------------------------------------------------------------------
%La trace actuelle est la trace \'emise par le traceur du processus observ\'e.
%%The actual trace is the trace emitted by a tracer, which can actually be ``visible''.
 
%The virtual full trace is a particular case of actual trace where the attributes values $A_t$ are the correspojnding parameter values $S_{t+1}$.
%La trace int\'egrale virtuelle est un cas particulier de trace actuelle o\`u les attributs d\'ecrivent compl\`etement les \'etats obtenus par la suite des transitions.

\begin{example}
In the Section~\ref{generation} we give a short example of actual trace, looking as follows (here the event identifier and the chrono are the same).

\begin{verbatim}
chrono attributes:
  1     1     1     Call    goal   
  2     2     2     Call    p(X)   
  3     2     2     Exit    p(a)   
  4     3     2     Call    eq(a,b) 
  5     3     2     Fail    eq(a,b)
  6     2     2     Redo    p(a)  
 ...
\end{verbatim}
\end{example}

It will be shown that it describes the evolution of a tree labeled by predications. However the labels corresponding to clauses are not described.
%The main question here is ``what can we  understand from the evolution of the full virtual trace just reading this actual one?''. Assume that the actual trace would be limited to 

If the actual trace would be limited to as illustrated below, the actual trace would just describe the evolution of a single tree without labels.

\begin{verbatim}
chrono attributes:
  1     1     Call
  2     2     Call
  3     2     Exit
  4     3     Call
  5     3     Fail
  6     2     Redo
 ...
\end{verbatim}
%one should understand far less from the observed process than with the previous trace. However we may understand something, depending from the knowledge we have of the semantics of the trace.

%\vspace{1mm}
There are thus two questions one has to consider. The first is related to the rebuilding function (the actual trace interpretation), i.e. how does one interpret the actual trace as (a part of) the full virtual trace; or what is described by the actual trace. The second concerns the existence of the extraction function.
The answer to the first question is given by the trace interpretation schema (description of the rebuilding function) which should be given with a trace. The second relates to the existence of a trace model, also called observational semantics.

%It is important to observe that a full virtual state may have many parameters which cannot be deduced from an actual trace. This is because the full virtual trace may represent several different processes.

%\vspace{2mm}

%----------------------------------------------------------------------------OS
%\section{Observational Semantics and Related Functions}
\section{Introducing the Observational Semantics}
\label{semobs}

We present a concrete representation of the observational semantics, the trace schema and the trace interpretation schema. Illustrative examples will be found in the forthcoming sections.

\vspace{1mm}
An Observational Semantics (OS) is a model of the full virtual trace production. It describes a transition function between full virtual states of ${\cal S}$ such that every transition gives rise to a unique trace event (virtual and actual).

%description of full virtual trace evolution without knowing the operational semantics of the process which produced it. 
In the case of a single observed process, the OS may be considered as a (likely partial) abstract model of the process. If several processes are to be considered with the same actual traces, the OS is thus an abstraction of the semantics of several procecesses.% which indeed can be considered as a semantics of (a part of) the trace.

%The Observational Semantics describes all possible full virtual traces as all possible sequences of states described by a finite number of parameters. It is defined by a set of all possible initial states and a state transition function. Each transition produces a trace event.

The transition function will be described by a finite set of rules $R$ (one rule by event type in the full virtual trace). It will be represented by a named fraction consisting of four components:

%\vspace{1mm} 
\begin{itemize}
\item A rule identifier (rule name).
\item A numerator with conditions on current values of the parameters identifying the subset of states to which the rule applies and with some intermediate computation.
\item A denominator describing the computations of the new values of the parameters (invariant parameters are not described).
\item External conditions (between braces) or properties relating parameters to elements not described by the parameters, but influencing the choice of the rules or the new values of the parameters.
\end{itemize}

%\vspace{1mm}
\noindent
% MODELE
\reglecontroleshort{Name}
{Conditions \,\,characterizing \,\,the \,\,current \,\,state}
{Computations \,\,of \,\,the \,\,new \,\,parameters}
{\{External \,\,Conditions\}}

\vspace{1mm} 
%Note that the distinction between elements in the numerator and in external position is arbitrary. However any expression using external information will be between braces\footnote{In this short introduction we do not formalize the notion of external condition. It is based on the idea that the evolution of some parameters of the virtual state may not be explicitely described in the OS.}.

%Each transition rule of the OS will be presented formally by a triple $<name, S, S'>$ where, by abuse of notation one will denote by $S$ the conditions caracterizing the current state $S_t$ to which the transition can be applied, and by $S'$ the resulting state  $S_{t+1}$ obtained by the given computations. 

\vspace{1mm}
To describe the OS, two kinds of functions will be used: those related to the described objects and their evolution in the virtual trace and those related to events or objects not described in this trace, but likely to occur in the observed process. The functions of the first category are known as ``auxiliary'', those of second kind ``external''. They relate to parameters not taken into account in the virtual trace. Finally one will also distinguish the functions exclusively used for computation of the attributes during the extraction of the trace, named ``auxiliary extraction functions'' and those exclusively used for the rebuilding named ``auxiliary rebuilding functions''.

\vspace{1mm}
The extraction function  ${\cal E}$ will be described by the same kind of rules whose denominator will contain only the produced trace event (attributes only). 
There is only one produced trace event per rule in the OS. Therefore
the extraction function consists of as many components as there are rules in $R$ and is denoted  ${\cal E} _r$ for each rule $r$, i.e. ${\cal E} = \{{\cal E}_r | r \in R \}$.

Each rule of the trace schema has the following form.

\vspace{1mm}
%Un sch\'ema de trace a la forme d\'ecrite \`a la figure~\ref{regextract}.
%\begin{figure}[h]\small
\noindent
\def\et{,\ \ }
% MODELE
\reglecontrole{Name}
{Computation  \,\, of \, the \,\, attributes}
{< unique \  trace \,\, event>}
%{}
{\{ External \,\, Conditions\}}

\vspace{1mm}
The rebuilding function may take several actual trace events as arguments. In the case of the box model two successive actual trace events and two attributes are sufficient to build a new actual state and to characterize the corresponding transition in the SO. In this case, it is possible to describe the rebuilding function as a familly of {\em local rebuilding functions} indexed by the same set $R$, ${\cal C} = \{{\cal C}_r | r \in R \}$. To describe it, the same kind of rule presentation is used.

\vspace{1mm}
\noindent
% MODELE
%\reglecontroletroisshort{Nom}
\reglecontrole{Name}
{ Rule \,\, identification \,\, conditions}
%{<Condition \,\, optionnelles \,\, d'extraction>}
{Rebuilding \,\, computations}
%{Elements \,\, on \,\, new \,\, virtual\,\,  state}
{\{ 
%\\ \phantom{xxxxxxxxxxxxxxxxxxxxxxxxxxxxxxxxxxxxxxxxxxx} 
Trace \,\, events \}}
%\,\, external\,\, conditions \}}
%\saut
%\caption{Forme des règles de reconstruction (schéma de reconstruction)}
%\label{regrecons}
%\end{figure}

\noindent
In this case, the actual trace is considered as an ``external'' information and therefore is given in the braces. The numerator of the rule contains the additional conditions which, together with the attributes of the trace events, are used to identify the corresponding applied rule of the OS.
% (condition of ``comprehension'').

The denominator contains the rebuilding computations (computation of the parameters of the actual state, starting from the given trace events and the previous actual state). The correspondence between the trace interpretation schema and the local rebuilding function is detailed in \cite{halchive}.

\vspace{1mm}
Notice that the three sets of rules (OS, trace schema and trace interpretation schema) are pairwise bijective.

\vspace{1mm}
The questions advocated at the end of the previous section concern the relationships between the observational semantics and the actual trace and its interpretation. Hence the notion of ``faithfulness''.

%The question now arises concerning the comprehension of a trace, i.e. the possibility of rebuilding a succession of possibly partial states, starting from a produced trace, without the direct recourse to the OS, but which corresponds, step by step, with the transitions of the OS which produced this trace. It is what the concept of faithful trace is for.

The faithfulness concerns actual states likely restricted to a subset of parameters. One notes $S/Q$ the restriction of a full state $S$ to the parameters $Q$. $Q$ will be called {\em current actual state} and $S/Q$ the virtual state restricted to the parameters of $Q$.% or, if there is no ambiguity, {\em current restricted virtual state}.

%%It will thus be supposed that the OS is described by a finite set of rules which constitutes a ``model of trace'', such that each rule produces a trace event. 
%The extraction function thus consists of as many components as there are rules in $R$ and is denoted  ${\cal E} _r$ for each rule $r$. In the same way one will use a {\em rebuilding function} $ {\cal C} _r$ described for as many components as there are rules and denoted $ {\cal C} _r$ for each rule $r$. The description of ${\cal C}$ is a {\em rebuilding schema}.

\begin{definition} [Faithful Trace Interpretation]

Given an actual state domain ${\cal Q}$, restriction of ${\cal S}$ to a subset of its parameters, a trace interpretation schema ${\cal C}$, an OS defined on ${\cal S}$ by a finite set of transitions $R$ and a trace schema ${\cal E}$,
%an actual trace $T_w = <Q_0, w^*_t> \forall  t \geq 0$ ($Q_0 = {S_0}/Q$), \\ $T_w $ is {\em faithful} for ${\cal Q}$ with regards to the full virtual trace $T_v = <S_0, v^*_t> \forall t \geq 0$, if
%if there exists a function $ {\cal F} $ such as

${\cal C}$ is a faithful trace interpretation w.r.t the OS and ${\cal E}$, if for all actual trace $T_w = <Q_0, w^*_t>$, $t \geq 0$:% such that, $\ {\cal C}(w^*_t,Q_0) = Q_t$:

\begin{quote}
%$\forall t \geq 0, \ {\cal C}(w^*_t,Q_0) = Q_t\ $ and

$\forall i \in [0 .. t-1], \ {\cal C}(w^*_i,Q_0) = S_i/Q \ \wedge$
%$\forall i \in [0 .. t-1], \ Q_i = S_i/Q \ \wedge$
%$\ \ \ \ \forall i \in [0 .. t-1], \ Q_i = S_i/Q \ \wedge $
$\exists r \in R, \ \ \, $such that$\,\  A_i = {\cal E}_r(S_i,S_{i+1})\,\,(A_i$ attributes of $w_i)$.
\end{quote}
\end{definition}

Faithfulness is a kind of total correctness property of an actual trace and the part of the virtual state evolution it represents, with regards to the trace model defined by the observational semantics and the extraction function.
It includes indeed some partial correctness statment wrt SO and ${\cal E}$ by the fact that every actual trace can be interpreted as (part of) a virtual trace produced by the SO and ${\cal E}$. It includes also a kind of completness statment (completeness of the observational semantics) since it also states that there is no other actual traces that the one produced by the SO and ${\cal E}$. In short, faithfulness expresses the commutativity of the extraction/rebuilding schemata.
%stipulates that in any actual trace, the sequence of actual states, which is a sequence of restricted virtual states, corresponds to a sequence of transitions of the OS from which the actual trace has been extracted.

%Faithfulness covers two aspects: correctness of the trace model (the actual states in any actual trace can be obtained from this model) and comprehensibility of the actual trace (the corresponding sequence of transitions in the OS can be deduced from a very restricted actual trace).
%A faithful trace contains all needed information to have a faithful partial view of the observed process, and, if the OS of the process is known, to rebuild the corresponding complete virtual full trace.

%==================================

\section{An Observational Semantics of the Byrd Box Model}
\label{soprologpur}

In his articles \cite{byrd80,byrd80ter}, Byrd illustrates his model using two schemata: a box with its famous four ``ports'' 
%(see the Figure~\ref{fig:boite}) 
and ``and/or-trees'', an already very widespread structure at that time, which combines representations of proof-tree and search-tree. It uses neither the concept of partial proof-tree, nor that of search-tree (SLD-tree), still little known, the seminal Clark's report \cite{clark79} having hardly just appeared.

%\begin{figure}[t]
%\begin{center}
%\includegraphics[width=0.5\linewidth]{Images/Box.eps}
%\end{center}
%\caption{Box Model as draw by Byrd \cite{byrd80}}
%\label{fig:boite}
%\end{figure}

Byrd fustigates nevertheless the implementors who, at backtrack (which is displayed in the trace by an event with port {\bf Redo}), force to return directly to the selected choice point, and do not express explicitely in the trace all the steps back. Byrd estimates that this is likely to lose the user and that it is preferable to demolish step by step what was explicitly made at the time of the successive uses of the clauses in order to solve goals.

Even if one wants to remain as close as possible to this model, we will however not follow this point of view, and will currently adopt that of the implementors, more widespread, and which seems quite as easy to understand, considering that the whole matter is formalized there. Indeed, the box model obliges to follow the calls of clauses through a system embedded boxes. It is thus easy to understand that, as each box has a unique identifier, the access to a choice-point, deeply located in a large box stacking, can be done as clearly by jumping directly to the deepest box rather than by descending carefuly the staircase resulting from stacking, or by following strictly the opposite way. One will thus avoid to explicitly detail the manner of reaching the box by backtracking.

Even if we do not describe exactly the model initially defined by Byrd, we estimate that we keep its historically essential elements, namely the building-visit of tree and the boxes in which clauses, or a subset of them, are stored. The approach formalized here will be refered as the  {\em simplified box model}.

\vspace{1mm}
The stacking of the boxes and its evolution will thus be described by a building-visit of tree in which each node corresponds to a box. The visit strategy corresponds to standard Prolog strategy (ISO Prolog \cite{alipie96}), that is to say a top-down left-to-right building-visit. Each new node, or box, is associated with a number which is incremented by 1 at each time a node is created.

Each node is labelled with a predication\footnote{According to the ISO Prolog glossary, a predication is a term whose outermost functor is a predicate.} and a packet of clauses. Each box is thus the root of a subtree which is spread as a ``treemap'', thus resulting in a kind of algebra of embedded boxes.

\vspace{1mm}
We use the  vocabulary of ISO Prolog \cite{alipie96}.

\vspace{2mm}
{\underline{\bf Full virtual trace parameters}}

\noindent
A full state has 9 parameters:

$\ \ \ \ \ \ \ \ \ \{ T , u , n, num, pred , claus , first, ct , flr\}$. 
\begin{enumerate}
\item {\bf $T$}: $T$ is a tree labelled by creation numbers, predications and subsets of clauses of the program $P$. It is described here by its functions of building-visit-rebuilding (see below) and its labels. No specific data representation is requested here. However we use in the examples a notation ``\`a la Dewey''. Each node is represented by a sequence of integers and for example denoted $\epsilon$, $1$, $11$, $12$, $112$, $\dots$. $\epsilon$ is the empty word, $1$, $11$ are direct successors and $11$, $12$ are ``brothers''. The lexicographic ordering is as follows: $u, v, w$ are words, $u < uv \ (v \not = \epsilon)$, and $uiv < ujw \ $if$ \ i < j$.

\item {\bf $u \in T$}:  $u$ is the current node in $T$ (visited box).

\item {\bf $n \in {\cal N}$}:  $n$ is a positive integer, number of the last created node in $T$.

\item {\bf $num: T \rightarrow {\cal N}$}. Abbrev.~: {\bf $nu$}.
 $nu(v)$ is the creation number (positive integer) labelling node $v$ in $T$. 

\item {\bf $pred: T \rightarrow {\cal H}$}. Abbrev.: {\bf $pd$}.
 $pd(v)$ is the predication labelling the node $v$ in $T$. It is a term of ${\cal H}$ (non ground Herbrand base).
 
\item {\bf $claus: T \rightarrow 2^P$}. Abbrev.~: {\bf $cl$}.
$cl(v)$ is the list of clauses in $P$ (same order as in $P$) whose heads use predication $pd(v)$ associated with node $v$ in $T$. $[]$ is the empty list.
Depending on the clauses  in $cl(v)$, one gets several models of trace. We will assume that only useful clauses are in $cl(v)$ (clauses whose head is unifiable with $pd(v)$). If $cl(v)$ is empty, there is no way to solve the corresponding predication $pd(v)$ and the node is thus in ``failure''. This list is created externally when a predication is called (see $claus\_pred\_init$ in external functions) and updated each time the node is revisited (see $update\_claus\_and\_pred$ in auxiliary functions). 

\item {\bf $first: T \rightarrow Bool$}. Abbrev.~: $fst$.
 $fst(v)$ is true iff $v$ is a not yet visited node in $T$ (it is a leaf of $T$). 

\item {\bf $ct \in Bool$}:  $ct = true$ iff all nodes in the tree $T$ have been completely visited (the current node is $\epsilon$ again after completion of a building-visit).
%and the building-visit is back to the root). 

\item {\bf $flr \in Bool$}:  $flr = true$ if (and not iff) the current subtree is failed. $flr = false$ otherwise (which does not mean that the subtree is successful)
\end{enumerate}

\vspace{2mm}
{\bf Initial state $S_0$}:

Due to space limits we will sometimes use $tu$ ($fa$) instead of $true$ (resp. $false$).

\vspace{1mm}

\noindent $\{ \{ \epsilon \} , \epsilon , 1, \{ (\epsilon, 1)\}, \{ (\epsilon, goal)\}, \{ (\epsilon, list\_of\_goal\_claus)\}, \{(\epsilon,tu)\}, fa, fa \}$

\vspace{1mm}
The model is based on a building-visit of partial proof-trees, built then rebuilt after backwards. The nodes are built just before being visited for the first time. Each visit of a node (or box) produces a trace event.

\vspace{1mm}
%\newpage
{\underline{\bf Auxiliary Functions}} (parameters manipulations): 
\begin{itemize}
\item {\bf $parent: T \rightarrow T$}. Abbrev.: {\bf $pt$}.
$pt(v)$ is the unique direct ancestor of $v$ in $T$. 
To simplify the model, it is assumed that $pt(\epsilon) = \epsilon$.

\item {\bf $leaf: T \rightarrow Bool$}. Abbrev.: {\bf $lf$}.
$lf(v)$ is true iff $v$ is a leaf of $T$.

%\item {\bf $has\_a\_brother: T \rightarrow Bool$ }: $has\_a\_brother(u)$ is true iff $u$ has a brother in $T$. ($pred(u)$ is not the last predication in the body of the currently used clause which is the first clause in the box of the parent node of $u$in $T$). The root ($\epsilon)$ has no brother. Abbrev.: {\bf $hab(u)$ }.
\item {\bf $may\_have\_new\_brother: T \rightarrow Bool$}. Abbrev.: {\bf $mhnb$}.
$mhnb(v)$ is true iff $pd(u)$ is not the last predication in the body of the currently used clause (which is the first clause in the box of the parent node of $v$ in $T$). The root (node $\epsilon$) has no brother.

%\item {\bf $create\_child: T \rightarrow T$ }: $v = create\_child(u)$ is the new child of $u$ in $T$. Abbrev.: {\bf $crc(u)$}.
\item {\bf $create\_child: T \rightarrow T$}. Abbrev.: {\bf $crc$}.
$crc(v)$ is the new child of $v$ in $T$.

\item {\bf $create\_new\_brother: T \rightarrow T$}. Abbrev.: {\bf $crnb$}.
$crnb(v)$ is the new brother of $v$ in $T$. Defined if $v$ is different from $\epsilon$.

\item {\bf $has\_a\_choice\_point: T \rightarrow Bool$}. Abbrev.: {\bf $hcp$}.
$hcp(v)$ is true iff there is a choice point $w$ in the subtree rooted $v$ in $T$ ($cl(w)$ has at least one clause).

\item {\bf $greatest\_choice\_point: T \rightarrow T$}.  Abbrev.: {\bf $gcp$}.
$w= gcp(v)$ is the greatest choice point in the subtree of root $v$ in $T$ (i.e. such that $cl(w)$ has at least one clause) according to the lexicographic ordering of nodes in $T$.

\item {\bf $fact: T \rightarrow Bool$}. Abbrev.: {\bf $ft$}.
$ft(v)$ is true iff the first clause in $cl(v)$ is a fact.

\item {\bf $update\_number: Fu,T \rightarrow Fu $}.  $Fu, T$ stands respectively for a set of functions and a set of nodes representing a tree. Abbrev.: {\bf $upn$}.
$upn(nu,v)$ updates the function $num$  removing all defining pairs related to removed nodes in $T$ until node $v$ (not removed).

\item {\bf $update\_claus\_and\_pred: F,T,{\cal H} \rightarrow F $}. Abbrev.: {\bf $upcp$}.
($F$ denotes one of the following functions) $upcp(claus, v)$, $upcp(pred,v)$ (2 arguments) or \linebreak $upcp(pred, v,p)$ 
(3 arguments): updates the functions $claus$ and $pred$ removing all pairs related to removed nodes in $T$ until node $v$, updating too, if requested by the external function $pred\_update$, the value of $pd(v)$ with the pair $(v,p)$ and updating the value of the function $claus$ at node $v$, removing the last used clause.
\end{itemize}

%\vspace{2mm}
%\newpage
{\underline{\bf External functions}}:

They correspond to the actions not described in the full virtual trace but with some influence, in particular all the aspects of the resolution related to the unification, which are omitted in this OS.
\begin{itemize}
\item {\bf $success: T \rightarrow Bool$}: Abbrev.: {\bf $scs$}. $scs(v)$ is true iff $v$ is a leaf in $T$ and the current predication has been successfuly unified with the head of the clause selected in the current box (box associated with the node $v$), or $v$ is not a leaf in $T$ and the subtree of root $v$ is a sub-proof-tree (all leaves are successful).

%le sous-arbre de racine $v$ de $T$ est un arbre de preuve (si 
%$v$ est une feuille et la pr\'edication courante a \'et\'e unifi\'ee avec succ\`es \ag la t\^ete de la clause choisie pour ce n\oe ud. 

\item {\bf $failure: T \rightarrow Bool$}. Abbrev.: {\bf$flr$}.
$flr(v)$ is true iff $v$ is a leaf and no head of any clause of the program can be unified with the current predication (in this case $cl(v)$ is empty).

%\item {\bf $claus\_pred\_init: T \rightarrow (pred, list\_of\_clauses)$}: $claus\_pred\_init(u)$ adds to the function $claus$ the pair $(u, list\_of\_clauses)$ where  $list\_of\_clauses$ is the list of clauses defining the predication $pred(u)$ which can be used successfuly to try all possible solutions (if empty, there is no solution for this predication), and $pred(u) = p$. Abbrev.: {\bf $cpini(u)$}.
\item {\bf $claus\_pred\_init: T \rightarrow (pred, list\_of\_clauses)$}. Abbrev.: {\bf $cpini$}. \linebreak
$(c,p) = cpini(v)$ (1) updates in the function $cl$ the pair $(v,c)$ where  $c$ is the list of clauses defining the predication $pd(v)$ which thus can be used successfuly to try all possible solutions, and (2) updates in the function $pd$ the pair $(v,p)$ where $p$ is the predication to be associated to the node $v$. The elements (clause and predication) computed by $cpini(v)$ will be respectively denoted $c\_cpini(v)$ and $p\_cpini(v)$.

\item {\bf $pred\_update: T \rightarrow {\cal H}$}. Abbrev.: {\bf $pud$}.
$pud(v)$ is the new value given to the current predication $pd(v)$ labelling the node $v$ in $T$, following a successful unification.
\end{itemize}

\begin{figure*}[ht]\small
\noindent
% CALL1
\reglecontrole{\callun{}}
{fst(u) \wedge lf(u) \wedge \neg ct \wedge ft(u) }
{cl' \gets upcp(cl,u) \et fst'(u) \gets fa \et flr' \gets fa }
{\{ \}}
\saut
% CALL2
\reglecontroledeux{\calldeux{}}
{fst(u) \wedge lf(u) \wedge \neg ct \wedge \neg ft(u) \et v \gets crc(u)}
%Modif Pierre le 20/4/07 car il n'y a pas besoin de propager l'effet de l'unification 
%{T' \gets T \cup  \{v\} \et u' \gets v \et n'\gets n+1 \et nu' \gets nu \cup \{(v,n')\} \et pd' \gets upcp(pd,u,p') \cup \{(v,p)\} \et }
{T' \gets T \cup  \{v\} \et u' \gets v \et n'\gets n+1 \et nu' \gets nu \cup \{(v,n')\} \et pd' \gets pd \cup \{(v,p)\} \et }
{ cl' \gets upcp(cl,u) \cup \{(v,c)\} \et fst'(u) \gets fa \et fst' \gets fst' \cup \{(v,tr)\} \et flr' \gets fa}
%{\{ \\ \phantom{xxxxxxxxxxx} scs(u) \et p'= pud(u) \et (c,p) = cpini(v)\}}
{\{ \\ \phantom{xxxxxxxxxxxxxxxxxxxxxxxxxxxxxxx} scs(u) \et (c,p) = cpini(v)\}}
\saut
% EXIT1
\reglecontrole{\exitun{}} 
{\neg fst(u) \wedge \neg mhnb(u) \wedge \neg ct \wedge \neg flr \et v \gets pt(u)}
{u' \gets v \et pd' \gets upcp(pd,u,p) \et (u = \epsilon) \Rightarrow (ct' \gets tr)}
{\{ scs(u) \et \\ \phantom{xxxxxxxxxxxxxxxxxxxxxxxxxxxxxxxxxxxxxxxxxxxxxxxxxxxxxx} p=pud(u)\}}
\saut
% EXIT2
\reglecontroledeux{\exitdeux{}}
{ \neg fst(u) \wedge mhnb(u) \wedge \neg ct \wedge \neg flr \et v \gets crnb(u)}
{T' \gets T \cup  \{v\} \et u' \gets v \et n'= n+1 \et nu' \gets nu \cup \{(v,n')\} \et}
{ pd' \gets upcp(pd,u,p') \cup \{(v,p)\} \et cl' \gets cl \cup \{(v,c)\} \et  fst' \gets fst \cup \{(v,tr)\}}
{ \{ \\
\phantom{xxxxxxxxxxxxxxxxxxxxxxxxxxxxxxxxx} scs(u) \et p' = pud(u) \et (c,p) = cpini(v) \}}
 \saut
% FAIL2
\reglecontrole{\faildeux{}}
{\neg fst(u) \wedge \neg ct \wedge \neg hcp(u) \et  v \gets pt(u)}
{u' \gets v \et (u = \epsilon) \Rightarrow (ct' \gets tr) \et flr' \gets tr}
{\{flr(u) \ \vee \  flr\}}
\saut
% REDO1
\reglecontroledeux{\redoun{}}
{ v \gets gcp(u) \et \neg fst(u) \wedge hcp(u) \wedge ft(v) \wedge (flr \ \vee \ ct)}
{T' \gets T - \{y | y > v \}\et u' \gets v \et cl' \gets upcp(cl,v) \et}
{ ct \Rightarrow (ct' \gets fa) \et flr' \gets fa}
{\{ \}}
 \saut
% REDO2
\reglecontroledeuxshort{\redodeuxshort{}}
%{ hcp(u) \wedge v = gcp(u) \wedge lf(v) \wedge \neg fact(v) \et w = crc(v)}
{ v \gets gcp(u) \et \neg fst(u) \wedge hcp(u) \wedge (flr \ \vee \ ct) \wedge \neg ft(v) \et w \gets crc(v)}
%{T' \gets T - \{y | y > v \} \cup \{w\} \et u' \gets w \et cl' \gets upcp(cl,v) \cup \{(w,c)\} \et pd' \gets pd \cup \{(w,p)\} \et ct \Rightarrow (ct \gets F)}
%{idem\  \redoun{} \ \et n'= n+1 \et nu' \gets upn(nu,v) \cup \{(w,n')\} \et   cl' \gets upcp(cl,v) \cup \{(w,c)\} \et pd' \gets pd \cup \{(w,p)\} \et ct \Rightarrow (ct \gets F)}
{ T' \gets T - \{y | y > v \} \cup \{w\}\et u' \gets w  \et n'= n+1 \et nu' \gets upn(nu,v) \cup \{(w,n')\} \et flr' \gets fa \et }
{pd' \gets upcp(pd,v) \cup \{(w,p)\}  \et cl' \gets upcp(cl,v) \cup \{(w,c)\} \et fst' \gets fst \cup \{(w,tr)\} \et ct' \Rightarrow (ct \gets fa)}
{\{ \\
\phantom{xxxxxxxxxxxxxxxxxxxxxxxxxxxxxxxxxxxxxxx}scs(v) \et (c,p) = cpini(w)\}}
% \saut
%\caption{Observational Semantics for Prolog (Virtual Trace)}
\caption{Observational Semantics of Prolog resolution (full virtual trace)}
\label{sovirttracefig}
\end{figure*}

Notice that $\forall u, flr(u) \Rightarrow flr = true$ (see the \faildeux{} rule).

\vspace{2mm} 
The OS is defined by the rules of the Figure~\ref{sovirttracefig}. Each rule is commented in the following.

\begin{itemize}
\item \callun{}: The current node is a leaf and the called predication will be solved by a fact. This node will thus remain a leaf. The choice point is updated (a clause is removed from the box).

\item \calldeux{}: the current node is a leaf but the associated predication is solved with a clause whose head has been successfuly unified and whose body is not empty. This node will be expanded. A new node is created of which box $v$ is filled with useful clauses and a calling predication is associated. The choice point is updated.

\item \exitun{}: successful exit from the last predication of the body of the current clause. $pred(u)$ is updated (it is not necessarily the same one as at the time of the call). Step up with a successful subtree without creation of any new branch.

\item \exitdeux{}: successful exit from a predication of the body of the current clause with creation a new ``sister'' (new leaf $v$, in case of using a clause with more than one predication in the body). The box $v$ is filled with useful clauses and node $v$ is labelled with a calling predication.

\item \faildeux{}: step up with a failed subtree as long as there is no choice point in the subtree.

\item \redoun{}: backtrack following success or failure, if there is a choice point in the subtree opening for a possible solution (or new solution if the current node is the root). As discussed at the beginning of this section, in this model, one does not repeat all the {\bf Redo}, following all the steps down and back until the choice point, as in the original Byrd's model.

\item \redodeux{}: backtrack following success or failure, if there is a choice point in the subtree, opening for a possible solution (or new solution if the current node is the root). As above, but with the creation of a successor as in the \calldeux{} rule.
\end{itemize}

%\vspace{2mm}
In the initial state $S_0$, only one of the rules \callun{} or \calldeux{} may apply. Whatever is the state, only one rule in $R$ can be applied as long as a complete tree has not been built. No rule applies any more if the built tree is complete and it does not contain choice point. 
%For the rules \callun{} and \calldeux{} the associated port in the trace is {\bf Call}, for the rules \exitun{} and \exitdeux{}, {\bf Exit}, for \faildeux{}, {\bf Fail}, and for \redoun{} \redodeux{}, the port is {\bf Redo}.

%===========================================
%fin traduction Pierre
\section{Extraction of the actual trace}
\label{generation}

Each application of a rule of the OS gives place to the extraction of a trace event  whose chrono is incremented of a unit each time. For the extraction one needs an auxiliary function.%; the chrono is not explicitely described.

\vspace{1mm}
%\newpage
{\underline{\bf Auxiliary extraction function 
}}.
\begin{itemize}
\item {\bf $lpath: T \rightarrow {\cal N}$}. Abbrev.: {\bf $lp$}. 

Byrd calls it the ``depth of recursion''. 
 $lp(v)$ 
is the number of nodes on the path from  the root to the node 
$v$.
It is thus the length of the path from the root to the node  $+ 1$. $lp(\epsilon) = 1$. 
%\item {\bf $rang\_node: T \rightarrow {\cal N}$ }: $rang\_nodes(u)$ est le rang du n\oe ud  $u$ dans $T$ dans l'ordre total de cr\'eation des n\oe uds. 
%%C'est donc, \'a une unit\'e pr\`es,le cardinal de $T$. 
%Ce nombre correspond au num\'ero attach\'e \`a une bo\^ite dans le mod\`ele de Byrd original. Abbrev.: {\bf $rgn(u)$}. $rgn(\epsilon)=1$.
%\item {\bf $a voir$ }Abbrev.: {\bf ... }. 
\end{itemize}

\vspace{2mm}
In the box model, the actual trace has 4 attributes and each event has the form
\begin{verbatim}
         t    r    l    port    p 
\end{verbatim}
where
\begin{itemize}
%\item \verb.id. est l'identificateur d'\'ev\'enement de trace.
\item \verb.t. is the  chrono.
\item \verb.r. is the creation number of the current node $u$ (the node in the current state), $nu(u)$.
\item \verb.l. is the depth in the tree $T$ of the current node, that is to say $lp(u)$.
\item \verb.port. is the action identifier having produced the trace event ({\bf Call, Exit, Fail} or {\bf Redo}).
\item \verb.p. is the predication associated with the current node, that is to say $pd(u)$.
\end{itemize}

\vspace{1mm}
%L'identificateur sera omis ici; c'est l'ordre de production des \'ev\'enements de trace, le chrono, qui en tiendra lieu.
%La trace actuelle ne contient que 4 types d'\'ev\'enements, mais aucune information particuli\`ere concernant quelle r\`egle exacte va s'appliquer. Comme il s'agit d'une trace int\'egrale, les \'el\'ements de chaque r\`egle doivent pouvoir \^etre retrouv\'es \`a partir des attributs de la trace, \`a savoir \`a partir des attributs $number(u)$, $lpath(u)$, et $pred(u)$.

Example 1 below presents a program and the extracted trace corresponding to the goal \verb.:- goal., ($u$  current node) 
\begin{verbatim}
c1: goal:-p(X),eq(X,b).
c2: p(a).
c3: p(b).
c4: eq(X,X).

:- goal. 
                    
chrono nu(u) lp(u)  port    pd(u)    Virtual State reached 
  1     1     1     Call    goal              S2
  2     2     2     Call    p(X)              S3
  3     2     2     Exit    p(a)              S4
  4     3     2     Call    eq(a,b)           S5
  5     3     2     Fail    eq(a,b)           S6
  6     2     2     Redo    p(a)              S7
  7     2     2     Exit    p(b)              S8
  8     4     2     Call    eq(b,b)           S9
  9     4     2     Exit    eq(b,b)           S10
 10     1     1     Exit    goal              S11
\end{verbatim}

%The example is detailed in the appendix B. 
%\vspace{2mm}
%La fonction d'extraction ${\cal E}$ 
The trace schema is described in figure\rf{tracegenfig}. Note that in these rules, the node $u$ refers to the current virtual state $S_t$.
\begin{figure*}[ht]\small
\noindent
%% CALL2
%\reglecontrolequatre{\calldeux{}}
%{lf(u) \wedge \neg ft(u) \wedge \neg ct \et v \gets crc(u)}
%{T' \gets T \cup  \{v\} \et u' \gets v \et n'= n+1 \et nu' \gets nu \cup \{(v,n')\} \et}
%{ cl' \gets upcp(cl,u) \cup \{(v,c)\} \et  pd' \gets pd \cup \{(v,p)\} \et flr' \gets false}
%{<nu(u) \ \ lp(u)\ \ {\bf CALL} \ \ pd(u)> }
%{\{ \\ \phantom{xxxxxxxxxxxxxxxxxxxxxxxxxxxxxxxxxxxxxxxxxxxxxxxxx} (c,p) = cpini(v)\}}
%\saut
%\saut
%%\end{figure}
%%\begin{figure}[ht]\small
%%\noindent
%% CALL1
\reglecontrole{\callun{}}
%{}
{}
{<nu(u) \ \ lp(u)\ \ {\bf Call} \ \ pd(u)>}
{\{ \}}
\saut
% CALL2
\reglecontrole{\calldeux{}}
%{}
{}
{<nu(u) \ \ lp(u)\ \ {\bf Call} \ \ pd(u)> }
{\{\}}
 \saut
% EXIT1
\reglecontrole{\exitun{}} 
%{}
{}
{<nu(u) \ \ lp(u)\ \ {\bf Exit} \ \ p>}
{\{p=pud(u)\}}
 \saut
% EXIT2
\reglecontrole{\exitdeux{}}
%{}
{}
{<nu(u) \ \ lp(u)\ \ {\bf Exit} \ \ p> }
{ \{ p = pud(u) \}}
 \saut
% FAIL2
\reglecontrole{\faildeux{}}
%{}
{}
{<nu(u) \ \ lp(u)\ \ {\bf Fail} \ \ pd(u)>}
{\{ \}}
\saut
% REDO1
\reglecontrole{\redoun{}}
{ v \gets gcp(u) }
%{ }
{<nu(v) \ \ lp(v)\ \ {\bf Redo} \ \ pd(v)> }
{\{ \}}
 \saut
% REDO2
\reglecontrole{\redodeux{}}
{ v \gets gcp(u) }
%{}
{ <nu(v) \ \ lp(v)\ \ {\bf Redo} \ \ pd(v) >}
{\{ \}}
\caption{Trace schema  (actual trace extraction function)}
\label{tracegenfig}
\end{figure*}

\vspace{2mm}
\section{Actual Trace Interpretation}
\label{reconstruction}

%As Byrd  trace,
%% (see Appendix A),
% in his original form\footnote{%A detailed description of the  original Byrd trace is given in appendices A and D. The corresponding original actual trace does not say anything on the evolution of the clauses themselves in the boxes. For this reason information on the clauses is omitted in the current actual state.}, only  describes the building-visit of tree whose nodes are labelled with predications  and  gives, in case of success, the complete final skeleton decorated with the final correct labels, one obtains thus finally  the instances  of clauses used, without necessarily knowing which clause was actually used at a given node.

To only account for the elements specific to the box model (evolution of the tree and predication labels) 4 parameters are sufficient. 
%ne donne pas suffisamment d'information pour permettre de reconna\ii tre \ag coup s\^ur les clauses choisies, on ne retient que 8 param\`etres. L'\'etat courant a donc  la forme: 
One will thus take as restricted virtual state the following parameters:  
\begin{quote}
$Q = \{ T , u , num, pred \}$.
%$\{ T , u , n, num, pred , first, ct , flr \}$. 
\end{quote}
%Il ne comporte plus que 8 param\`etres. 
%On ne d\'ecrit pas non plus l'\'evolution du programme $P$ car la trace ne contient pas non plus d'information sur son \'evolution.
Note that one could have added the parameters 
$ct$ and $flr$. But that does not appear necessary a priori because $ct$ is true 
(except at the first trace event) iff the first or the second attribute is $1$ (in fact   they are $1$ together); and $flr$ becomes false (failure) for any trace event  of port {\bf Fail}. 
In particular if we are at the root, we know  if we are in failure (event of port {\bf Fail} at the root) or in success (event of port {\bf Exit} at the root).
 Then we know   if we have a failure  tree  or a complete proof tree  (success). 

\vspace{2mm}

So the initial state $S_0/Q$ is: (see the complete state at the previous section)%{\bf Etat initial}: 
\begin{quote}
\noindent $\{ \{ \epsilon \} , \epsilon , \{ (\epsilon, 1)\}, \{ (\epsilon, goal)\}\}$
%\noindent $\{ \{ \epsilon \} , \epsilon , 1, \{ (\epsilon, 1)\}, \{ (\epsilon, goal)\}, \{ (\epsilon, true)\}, false, false\}$
\end{quote}
%Le principe est simple: il faut  retrouver la r\`egle de la SO qui s'applique \`a chaque pas de trace avec pour seule info externe cette fois les infos donn\'ees dans la trace. D'o\`u l'allure (voir premi\`ere r\`egle).

%Pour y arriver, il suffit d'observer que les deux premiers nombres fournis permettent de construire un arbre. La trace \'emise permet donc de construire l'arbre et, \`a partir de l\`a de conna\^itre la r\`egle de la SO qui s'applique. On arrive alors \`a reconstituer en partir la trace virtuelle. On verra qu'on ne la reconstitue qu'en partie et que donc la trace actuelle n'est pas ``int\'egrale'' dans ce mod\`le.

The  rebuilding function  ${\cal C}$ of the restricted virtual trace
now is described starting from the initial actual state and the actual trace.
%as well as the faithfulness of the actual trace for this state relative with the virtual trace. 
%La figure~\ref{traceregenefig} donne les r\`egles de reconstruction relativement \`a un \'etat courant dont l'\'evolution des clauses a \'et\'e supprim\'ee. 

\vspace{1mm}
{\underline{\bf Auxiliary function of rebuilding }}:

To rebuild the partial current state, one auxiliary function only is necessary, namely the inverse function of $num$, noted $node$. 
%qui associe au premier attribut de la trace $r$ -un entier positif ici- un n\oe ud dans l'arbre courant $T$ (celui dont l'ordre de cr\'eation est $r$, voir section~\ref{semobs}).
\begin{itemize}
\item {\bf $node: {\cal N} \rightarrow T$}. Abbrev.: {\bf $nd$}.
 Inverse function of  $num$. $v= nd(n)$ is the node of  $T$ whose creation rank is  $n$ (such that $nu(v) = n$). By definition $nd(nu(v))=v$ and $nu(nd(n))=n$.

%\item {\bf $rebuild\_node: {\cal N},{\cal N} \rightarrow T$ }:  c'est la fonction inverse de $number$ et $lpath$ ensembles; $u = rebuild\_node(r,l)$  est le n\oe ud $u$ dans $T$ dont le num\'ero est $r$ et la profondeur $l$. Pour un arbre donn\'e, c'est une bijection. Abbrev. $rbn(r,l)$.
%\item {\bf $node: {\cal N} \rightarrow T$ }: $u = nodes(n)$ est le n\oe uds de num\'ero  $n$ in $T$. Abbrev.: {\bf $nd(u)$}.
%\item {\bf $a voir$ }Abbrev.: {\bf ... }.  
\end{itemize}

\vspace{2mm}
The interpretation schema is given in figure~\ref{traceregenefig} by the family $\{{\cal C}_r | r \in R \}$.

%Each rule comprises in numerator the condition of identification of the rule starting from the trace, in denominator calculations of the new virtual  restricted state starting from the trace events  which appear between braces and, possibly, the parameters of the current restricted virtual state. 
%They express that 

%If $Cond_r(e_t,e'_{t+1}) \ $ then ${\cal C}_r(e_t,e'_{t+1},Q_t) = Q_{t+1}$.

%These rules make it possible in particular to rebuild a tree step by step (or in an equivalent way embedded boxes), its building-rebuilding-visit, and the functions $num$ and $pred$. 
%The restricted virtual state thus comprises 4 parameters, namely 
% $Q = S/Q = \{ T, u, num, pred \}$
%et d'obtenir dans l'\'etat final ($S_{11}$) un arbre de preuve complet.
%Reconstruire la fonction $clause$ est beaucoup plus probl\'matique comme on le verra \`a la fin.

\begin{figure*}[ht]\small
\noindent
%$\{ T ,            u ,     (n),  nu,             pd ,                 (fst),            (ct) , (flr)\}$
%$\{\{\epsilon \}, \epsilon, 1, \{(\epsilon,1)\}, \{(\epsilon,goal)\}, \{(\epsilon,true)\},false,false\}$
% CALL1
\reglecontrole{\callun{}}
{r' = r}
{}
%{}
{\{< r \ \ l \ \ {\bf Call} \ \ p > \pv < r' >\}}
\saut
%% CALL1
%\reglecontrole{\callun{}}
%{fst(u) \wedge lf(u) \wedge \neg ct \wedge ft(u) }
%{cl' \gets upcp(cl,u) \et fst'(u) \gets false \et flr' \gets false}
%{\{ \}}
%\saut
%$\{ T ,            u ,     (n),  nu,             pd ,                 (fst),             (ct), (flr)\}$
%$\{\{\epsilon \}, \epsilon, 1, \{(\epsilon,1)\}, \{(\epsilon,goal)\}, \{(\epsilon,true)\},false,false\}$
% CALL2
\reglecontrole{\calldeux{}}
{r' > r}
%{}
{u' \gets crc(nd(r)) \et T' \gets T \cup  \{u'\} \et nu'(u') \gets r' \et pd'(u') \gets p' }
%{}
{\{\\ \phantom{xxxxxxxxxxxxxxxxxxxxxxxxxxxxxxxxxxxxxxxxxxx} 
< r \ \ l \ \ {\bf Call} \ \ p > \pv <r' p'>\}}
 \saut
%% CALL2
%\reglecontroledeux{\calldeux{}}
%{fst(u) \wedge lf(u) \wedge \neg ct \wedge \neg ft(u) \et v \gets crc(u)}
%{T' \gets T \cup  \{v\} \et u' \gets v \et n'\gets n+1 \et nu' \gets nu \cup \{(v,n')\} \et pd' \gets upcp(pd,u,p') \cup \{(v,p)\} \et }
%{ cl' \gets upcp(cl,u) \cup \{(v,c)\} \et fst'(u) \gets false \et fst' \gets fst' \cup \{(v,true)\} \et flr' \gets false}
%{\{ \\ \phantom{xxxxxxxxxxx} scs(u) \et p'= pud(u) \et (c,p) = cpini(v)\}}
% \saut
%$\{ T ,            u ,     (n),  nu,             pd ,                 (fst),             (ct), (flr)\}$
%$\{\{\epsilon \}, \epsilon, 1, \{(\epsilon,1)\}, \{(\epsilon,goal)\}, \{(\epsilon,true)\},false,false\}$
% EXIT1
\reglecontrole{\exitun{}} 
{r' < r \vee u = \epsilon}
%{}
{u' \gets pt(u) \et pd'(u) \gets p}
{\{ < r \ \ l \ \ {\bf Exit} \ \ p > \pv < r' > \}}
 \saut
%% EXIT1
%\reglecontrole{\exitun{}} 
%{\neg fst(u) \wedge \neg mhnb(u) \wedge \neg ct \wedge \neg flr \et v \gets pt(u)}
%{u' \gets v \et pd' \gets upcp(pd,u,p) \et (u = \epsilon) \Rightarrow (ct' \gets true)}
%{\{ scs(u) \et \\ \phantom{xxxxxxxxxxxxxxxxxxxxxxxxxxxxxxxxxxxxxxxxxxxxxxxxxxxxxx} p=pud(u)\}}
% \saut
%$\{ T ,            u ,     (n),  nu,             pd ,                 (fst),             (ct), (flr)\}$
%$\{\{\epsilon \}, \epsilon, 1, \{(\epsilon,1)\}, \{(\epsilon,goal)\}, \{(\epsilon,true)\},false,false\}$
% EXIT4
\reglecontroleshort{\exitdeuxshort{}}
{  r' > r \wedge u \not = \epsilon}
%{}
%{v \gets nd(r) \et u \gets crnb(v) \et T' \gets T \cup  \{u\} \et nu(u) \gets r' \et pd(v) \gets p \et pd(u) \gets p'}
{u' \gets crnb(u) \et T' \gets T \cup  \{u'\} \et nu'(u') \gets r' \et pd'(u) \gets p \et pd'(u') \gets p'}
{\{ \\ \phantom{xxxxxxxxxxxxxxxxxxxxxxxxxxxxxxxxxxxx}
< r \ \ l \ \ {\bf Exit} \ \ p > \pv <r' \ \ p'> \}}
 \saut
%% EXIT2
%\reglecontroledeux{\exitdeux{}}
%{ \neg fst(u) \wedge mhnb(u) \wedge \neg ct \wedge \neg flr \et v \gets crnb(u)}
%{T' \gets T \cup  \{v\} \et u' \gets v \et n'= n+1 \et nu' \gets nu \cup \{(v,n')\} \et}
%{ pd' \gets upcp(pd,u,p') \cup \{(v,p)\} \et cl' \gets cl \cup \{(v,c)\} \et  fst' \gets fst \cup \{(v,true)\}}
%{ \{ \\
%\phantom{xxxxxxxxxxxxxxxxxxxxxxxxxxxxxxxxx} scs(u) \et p' = pud(u) \et (c,p) = cpini(v) \}}
% \saut
%$\{ T ,            u ,     (n),  nu,             pd ,                 (fst),             (ct), (flr)\}$
%$\{\{\epsilon \}, \epsilon, 1, \{(\epsilon,1)\}, \{(\epsilon,goal)\}, \{(\epsilon,true)\},false,false\}$
% FAIL2
\reglecontrole{\faildeux{}} 
{\,\,\,\,\,\,\,\,\,\,\,\,}
%{<un(r) = u \wedge pd(u) = p>>}
{u' \gets pt(u)}
{\{< r \ \ l \ \ {\bf Fail} \ \ p > \}}
 \saut
%% FAIL2
%\reglecontrole{\faildeux{}}
%{\neg fst(u) \wedge \neg ct \wedge \neg hcp(u) \et  v \gets pt(u)}
%{u' \gets v \et (u = \epsilon) \Rightarrow (ct' \gets true) \et flr' \gets true}
%{\{flr(u) \ \vee \  flr\}}
%\saut
%$\{ T ,            u ,     (n),  nu,             pd ,                 (fst),             (ct), (flr)\}$
%$\{\{\epsilon \}, \epsilon, 1, \{(\epsilon,1)\}, \{(\epsilon,goal)\}, \{(\epsilon,true)\},false,false\}$
%REDO1
\reglecontrole{\redoun{}}
{r' = r}
%{< v = un(r) \et \wedge pd(v) = p >}
{u' \gets nd(r) \et T' \gets T - \{y | y > u'\} }
{\{ < r \ \ l \ \ {\bf Redo} \ \ p > \pv < r' >\}}
 \saut
% REDO1
%\reglecontroledeux{\redoun{}}
%{ v \gets gcp(u) \et hcp(u) \wedge ft(v) \wedge (flr \ \vee \ ct)}
%{T' \gets T - \{y | y > v \} \et u' \gets v \et cl' \gets upcp(cl,v) \et}
%{ ct \Rightarrow (ct' \gets false) \et flr' \gets false}
%{\{ \}}
% \saut
%$\{ T ,            u ,     (n),  nu,             pd ,                 (fst),             (ct), (flr)\}$
%$\{\{\epsilon \}, \epsilon, 1, \{(\epsilon,1)\}, \{(\epsilon,goal)\}, \{(\epsilon,true)\},false,false\}$
%REDO2
\reglecontroledeuxshort{\redodeuxshort{}}
{r' > r }
%{<v = un(r) \wedge pd(v) = p \wedge r' = n+1>}
{v \gets nd(r) \et T' \gets T - \{y | y > v\}\cup \{u'\} \et u' \gets crc(v) \et}
{ nu' \gets upn(nu,v)\cup\{(u',r')\} \et pd' \gets upcp(pd,v)\cup\{(u',p')\}}
%{ nu' \gets upn(nu,v) \cup \{(w,n')\} \} }
{\{ 
\\ \phantom{xxxxxxxxxxxxxxxxxxxxxxxxxxxxxxxxxxxx}
 < r \ \ l \ \ {\bf Redo} \ \ p > \pv < r' \ \ p' >\}}
 \saut
% REDO2
%\reglecontroledeux{\redodeux{}}
%{ v \gets gcp(u) \et hcp(u) \wedge (flr \ \vee \ ct) \wedge \neg ft(v) \et w \gets crc(v)}
%{ T' \gets T - \{y | y > v \} \et u' \gets w  \et n'= n+1 \et nu' \gets upn(nu,v) \cup \{(w,n')\} \et pd' \gets pd \cup \{(w,p)\}  \et}
%{cl' \gets upcp(cl,v) \cup \{(w,c)\} \et fst' \gets fst \cup \{(w,true)\} \et ct' \Rightarrow (ct \gets false)\et flr' \gets false}
%{\{ \\
%\phantom{xxxxxxxxxxxxxxxxxxxxxxxxxxxxxxxxxxxxxxxxxxxxxxxx}(c,p) = cpini(w)\}}
\caption{
Trace Interpretation Schema (simplified box model)}
\label{traceregenefig}
\end{figure*}

\vspace{1mm}
%To understand this trace requires to read two successive trace events. 
 %Par ailleurs et afin de faciliter la lecture, les informations que l'on peut reprendre des r\`egles de la SO sont omises. 

%\vspace{2mm}
%It  should be noticed that  when the trace is faithful, one can thus reconstitute the ``operating'' of the OS starting from the trace,
%****suite supprimee

\vspace{1mm}
A complete proof of the faithfulness of the interpretation trace schema ${\cal C}$ wrt the given OS and ${\cal E}$, can be found \cite{halchive}.
%s given in  appendix C.%------------------------------sssssss repris dans versions finale jfpc yyyyyyyyyyyyyyyy

%============================================================
\section{Discussion and Conclusion}
\label{commdisc}

The main point of this paper is the illustration of an original approach to give a semantics to traces. % distincte d'une simple abstraction du processus trac\'e. 
The example of the box model, used here, is primarily anecdotic. But, in fine, the result is undoubtedly a complete formalization, which is also one of  the simplest formalizations of this model (simple because restricted to only elements necessary to its comprehension).

Our first observations will relate to the comprehension of the trace given by the rules of figure~\ref{traceregenefig}.
They give an first interpretation of the trace, provided the actual trace reflects a true execution (for example it is the trace of a tree construction according to some strategy). The faithfulness property guarantees that it can be related with a more complete semantics given by the observational semantics and the associated actual trace extraction function.
 
%Those can be understood indeed being limited in a restricted state (denoted $Q$) without having recourse to the complete OS. All that is necessary there is formalized, the recourse to the OS being useful to only look further into comprehension. 
%The rules give its dynamic skeleton  (building-rebuilding-visit of tree), and their associated optional conditions (always valid for the rebuilding of an actual trace produced with the OS) give the immediate interpretation of the attributes of the trace. 

This approach also immediately highlights the difficulties of interpretation of such a model. We will retain two of them. 
%En premi\`ere remarque, on observe 
At first it will be observed that the trace interpretation requires to analyse two actual trace events (Section~\ref{reconstruction}).
%, if it is normal that the interpretation of the trace requires to apprehend the whole of the trace since the beginning (to have an idea of the state of the resolution), it is less normal than the reading of an event ``ahead'' is necessary, which is  a factor of difficulty.
This could be avoided if some information on the clause would be displayed in the actual trace event (fact or not fact), making thus the trace simpler to read.
%\footnote{Indeed, under the conditions, the factors discriminating the used rules are  related to the nature of the clauses.} (for example the selected clause before an event of port {\bf Call}).
% The representation with boxes aimed primarily ``to contain'' the potentially useful clauses. This clearly make more difficult to understand 
%They are not found in the trace, this withdraws in the model most of its interest, by limiting it in fact to the only description of a  tree traversal. 
%ce qui retire au mod\`ele une  partie de son int\'er\^et en le limitant en fait \`a la seule description du parcours-construction-reconstruction d'un arbre de preuve.

In second remark one will observe a contrario that the trace contains an useless attribute. 
The depth (attribute {\tt l}) finally does not contribute to the comprehension of the trace and overloads it unnecessarily. 
In fact the depth could contribute to the comprehension of the partial proof tree  by combining it with an adequate coding of the nodes. 
This choice is made for example in the trace of \gprolog\ \cite{gnuprolog}
where the nodes are coded, not by their order of creation, but by their rank in the tree. 
The combination of the two attributes then allows a direct location of the current node in the tree $T$. 
This choice constitutes indeed an improvement of the original trace.
%\footnote{Many work introduce visualizations of the trace with indentation and use the attribute $lpath$ with this intention. That shows that this attribute has a practical utility, but it is not useful for the rebuilding.}.

\vspace{1mm}
The few articles quoted in the introduction translate the permanent search for improvements of the comprehension of control and also of the unification. 
So \cite{boiz84} (1984) \cite{NumFuji85} (1985) 
propose improvements of  Byrd trace with a more reduced number of events, thus bringing a more synthetic vision of the traversed tree, 
and they also propose new ports concerning the unification and the choice of the clauses. 
\cite{ToBe93} (1993)
explicitly introduces an algebra of boxes supported by  graphics, but this model, which wants to deal with all the aspects of the resolution, remains rather complex. 
 \cite{jahier00} (2000) 
proposes a semantics of trace based on a denotational semantics of Prolog. 
The principal disadvantage is that the comprehension of the trace needs a good comprehension of a complete model of Prolog, synthetic but requiring a certain familiarity with the continuations. 
In the paper \cite{kulas03} (2003) the approach is similar
but it is based directly on the ports whose possible sequences constitute its skeleton. 
The result is also that the comprehension of the trace needs the assimilation of a relatively complex semantics of Prolog, which is connected more with a semantics based on the ``magic sets'' than with a direct explanation of the trace.

\vspace{1mm} 
One may consider that the trace model presented here is not so easy to understand, since some auxiliary and external functions are not formally described. However the auxiliary functions always refer to wellknown data structure manipulation with an usual unambiguous semantics. Some external ones refer to a non described Prolog semantics (successful unification, clauses selection, proof-tree, ...) and are not defined here. It must be clear that the given model is a formal definition of a tracer, not of the observed processor. It means that the given model is not an executable specification and that its semantics depends on the interpretation given to undefined functions. In particular the faithfulness property does not depend on these interpretations. The interest and potential simplicity of such an approach precisely lies in the fact that there is no need to describe entirely the semantics of the observed processes.
%\vspace{1mm}
%These studies show that much was done to improve the means of  understanding  resolution. 
%In the course of time, work concentrated on methods of increasingly complex analysis and visualization (for example \cite{Opium:JLP} for analysis of  Prolog traces) for forms of resolution  also increasingly complex like the resolution of CSP \cite{oadymppac}. 
%However  Byrd trace remains the basis of the tracers for the systems of resolution and its famous ports still inspire, from time to time, the researchers. 

%%\cite{shapiro83},
%In this example one treated a particular instance of the box model. 
%It would be interesting, and it will be a next work, to obtain a more generic model which could  potentially generate various known implementations of this model. That is possible with the approach presented here. 
%%A first description of various models is made in  appendix D, with a OS proposed in appendix E. 

\nopagebreak

% R\'ef\'erences bibliographiques
%\bibliographystyle{plain}
\bibliographystyle{splncs}
%\nocite{*}
%nocite met tt la biblio

%\bibliography{toutenun}
%\bibliography{toutenun,pierre}
\bibliography{wlpe07corr}

\begin{thebibliography}{10}

\bibitem{ercimlnai}
Langevine, L., Deransart, P., Ducass\'e, M.:
\newblock A generic trace schema for the portability of cp(fd) debugging tools.
\newblock In Apt, K., Fages, F., Rossi, F., Szeredi, P., Vancza, J., eds.:
  Recent Advances in Constraints, 2003. Number 3010 in LNAI.
\newblock Springer Verlag (2004)

\bibitem{byrd80}
Byrd, L.:
\newblock Understanding the control flow of {Prolog} programs.
\newblock In Tarnlund, S.A., ed.: Logic Programming Workshop, Debrecen, Hungary
  (1980)

\bibitem{proedinb}
Moniz-Pereira, L., Pereira, F., Warren, D.:
\newblock {User's Guide to DECsystem-10 Prolog} (1978) {University of
  Edinburgh}.

\bibitem{promars}
Roussel, P.:
\newblock {Prolog : Manuel de R\'ef\'erence et d'Utilisation} (1975)
  Universit\'e d'Aix-Marseille II.

\bibitem{boiz84}
Boizumault, P.:
\newblock {Deux Mod\`eles de Trace pour le Langage Prolog}.
\newblock In Dincbas, M., Bourgault, S., eds.: Actes du Quatri\`eme S\'eminaire
  de Programmation en Logique, CNET-Lannion (France) (1984)

\bibitem{ToBe93}
Toberman, G., Berckstein, C.:
\newblock {What's in a Trace: The Box Model revisited.}
\newblock In Fritzson, P., ed.: Proceedings of the First Workshop on Automated
  and Algorithmic Debugging (AADEGUG'93). Number 749 in LNCS, Linkoeping,
  Sweden (1993)

\bibitem{jahier00}
Jahier, E., Ducass\'{e}, M., Ridoux, O.:
\newblock Specifying {Prolog} trace models with a continuation semantics.
\newblock In Lau, K.K., ed.: Proc. of {LO}gic-based Program Synthesis and
  TRansformation, London, Springer-Verlag, LNCS 2042 (2000)

\bibitem{kulas03}
Kul\`as, M.:
\newblock {Pure Prolog Execution in 21 Rules}.
\newblock In {Arnaud Lallouet}, ed.: Proc. of the 5th Workshop on Rule-Based
  Constraint Reasoning and Programming (RCoRP'03), Kinsale (2003) {Repository
  arXiv:cs:PL/0310020 v1}.

\bibitem{halchive}
Deransart, P., Ducass\'e, M., Ferrand, G.:
\newblock {Une s\'emantique observationnelle du mod\`ele des bo\^ites pour la
  r\'esolution de programmes logiques}.
\newblock Technical report, INRIA (2007) http://hal.inria.fr/inria-00151285.

\bibitem{pierrewlpe06}
Deransart, P.:
\newblock {On using Tracer Driver for External Dynamic Process Observation}.
\newblock In W., V., Mu\`noz-Hernandez, S., eds.: Proceedings of the 16th
  Workshop on Logic-based Methods in Programming Environments ({WLPE'06}), a
  pre-conference workshop of {ICLP}'06, Seattle, USA (2006) {\tt
  http://arxiv.org/abs/cs/0701148}.

\bibitem{byrd80ter}
Byrd, L.:
\newblock {Prolog} debugging facilities.
\newblock Technical Report D.A.I. paper No 19, University of Edinburgh (1980)

\bibitem{clark79}
Clark, K.:
\newblock {Predicate Logic as a Computational Formalism}.
\newblock Technical Report 79/59, Imperial College, London (1979)

\bibitem{alipie96}
Deransart, P., Ed-Dbali, A., Cervoni, L.:
\newblock {{Prolog}, The Standard; Reference Manual}.
\newblock Springer Verlag (1996)

\bibitem{gnuprolog}
Diaz, D.:
\newblock {{\sc GNU}-Prolog}, a free {Prolog} compiler with constraint solving
  over finite domains (2003) {\tt http://gprolog.sourceforge.net/}, Distributed
  under the GNU license.

\bibitem{NumFuji85}
Numao, M., Fujisaki, T.:
\newblock {Visual Debuggger for Prolog}.
\newblock In: Proceedings of the Second Conference on Artificial Intelligence
  Applications, Miami (1985)

\end{thebibliography}

\end{document}